\documentclass{emulateapj}
\usepackage{color}
\usepackage{bm}
\slugcomment{\scriptsize{Accepted for Publication in ApJ Letter, Preprint typeset using \LaTeX style}} 
\shorttitle{MHD Turbulence Powered by MRI in PNSs}
\shortauthors{Masada et al.}
\begin{document}
%%%%%%%%%%%%%%%%%%%%%%%%%%%%%%%%%%%%%%%%%%%%%%%%%%%%%%%%%%%%%%%%%%%%
%%%%%%%%%%----------------------------------- Title  & Authors  ---------------------------------%%%%%%%%%%%%%%%%%%%%%%%%
%%%%%%%%%%%%%%%%%%%%%%%%%%%%%%%%%%%%%%%%%%%%%%%%%%%%%%%%%%%%%%%%%%%%
\title{Magnetohydrodynamic Turbulence Powered by \\ Magnetorotational Instability in Nascent Proto-Neutron Stars}
\author{Youhei MASADA\altaffilmark{1}, Tomoya TAKIWAKI\altaffilmark{2} and Kei KOTAKE\altaffilmark{3}} 
\altaffiltext{1}{Department of Computational Science, Kobe University; Kobe 657-8501, Japan: E-mail: ymasada@harbor.kobe-u.ac.jp}
\altaffiltext{2}{Division of Theoretical Astronomy, National Astronomical Observatory of Japan; Tokyo 181-8588}
\altaffiltext{3}{Faculty of Science, Department of Applied Physics, Fukuoka University; Fukuoka 814-0180}
%%%%%%%%%%%%%%%%%%%%%%%%%%%%%%%%%%%%%%%%%%%%%%%%%%%%%%%%%%%%%%%%%%%%
%%%%%%%%%%%----------------------------------- Abstract --------------------------------------%%%%%%%%%%%%%%%%%%%%%%%%%
%%%%%%%%%%%%%%%%%%%%%%%%%%%%%%%%%%%%%%%%%%%%%%%%%%%%%%%%%%%%%%%%%%%%
\begin{abstract}
Magnetorotational instability (MRI) in a convectively-stable layer around the neutrinosphere is simulated by a three-dimensional model of 
supernova core. To resolve MRI-unstable modes, a thin layer approximation considering only the radial global stratification is adopted. 
Our intriguing finding is that the convectively-stable layer around the neutrinosphere becomes fully-turbulent due to the MRI and its 
nonlinear penetration into the strongly-stratified MRI-stable region. The intensity of the MRI-driven turbulence increases with magnetic flux 
threading the core, but is limited by a free energy stored in the differential rotation. The turbulent neutrinosphere is a natural consequence 
of rotating core-collapse and could exert a positive impact on the supernova mechanism. 
\end{abstract}
%%%%%%%%%%%%%%%%%%%%%%%%%%%%%%%%%%%%%%%%%%%%%%%%%%%%%%%%%%%%%%%%%%%%
%%%%%%%%%%%%%%%----------------------- Body of Paper --------------------%%%%%%%%%%%%%%%%%%%%%%%%%%%%%
%%%%%%%%%%%%%%%%%%%%%%%%%%%%%%%%%%%%%%%%%%%%%%%%%%%%%%%%%%%%%%%%%%%%
%%%%%%%%%%%%%%%-----------------------S1 Introduction------------------%%%%%%%%%%%%%%%%%%%%%%%%%%%%%%
%%%%%%%%%%%%%%%%%%%%%%%%%%%%%%%%%%%%%%%%%%%%%%%%%%%%%%%%%%%%%%%%%%%%
\section{Introduction}
%%%%%%%%%%%%%%%%%%%%%%%%%%%%%%%%%%%%%%%%%%%%%%%%%%%%%%%%%%%%%%%%%%%%
The magnetic field is not an exotic fuel for core-collapse supernovae (CCSNe) because it has self-excited and self-sustained 
natures and is an inevitable outcome of electrically-conducting fluid motions \citep[e.g.,][]{moffatt78}. Once it is generated, it cannot be 
dissipated during CCSNe due to its long dissipation time ($\simeq 10^{15}\ {\rm sec}$) \citep[e.g.,][]{masada+07}. The mechanism of 
CCSNe should be thus studied self-consistently in the framework of magnetohydrodynamics (MHD)
\citep[e.g.,][]{obergaulinger+06, burrows+07, takiwaki+11,winteler+12, endeve+12, moesta+14}. See \citet{kotake+12} and references 
therein for the MHD effect on the CCSNe. 

The higher the magnetic field strength and the pre-collapse rotation rate, the more MHD effect becomes important in the supernova 
dynamics. At the post-collapse stage, the magnetic field is amplified by two processes: one is the field wrapping (i.e., $\Omega$-effect), 
and the other is so-called magnetorotational instability (MRI, see \citealt{balb98}). The later is highlighted here because it can amplify 
the magnetic field exponentially in the differentially rotating core, much faster than the linear amplification due to the $\Omega$-effect. 
 
\citet{akiyama+03} were the first to point out that nascent protoneutron stars (PNSs) is generally subjected to the MRI. They predicted 
that, due to the MRI, the magnetic fields of $\sim 10^{15}$--$10^{16}$ G can develop after the core-bounce, which is high enough to 
affect the supernova dynamics. However, the nonlinear properties of the MRI in the CCSNe has not been fully elucidated. The short 
wavelength of the MRI prevents from accurately capturing it in the global simulation. 

To overcome this difficulty, local models are often used. \citet{obergaulinger+09} reported local shearing-disk simulations to study the 
nonlinear development of the MRI in the supernova core, and showed that the MRI amplifies the seed fields exceeding $10^{15}$ G  
as estimated in \citet{akiyama+03}. By local shearing box calculations of a small patch of the supernova core, Masada et al. (2012) 
presented a sub-grid scale model of the MRI-driven turbulence to predict the saturation amplitudes of the magnetic energy and turbulent 
stress. They also showed that the turbulent heating sustained by the MRI might play a crucial role in facilitating the neutrino-heating 
mechanism \citep[see also][]{thompson+05}. 

Only by dropping the dimension, the MRI can be simulated even in the global model. A pioneering work by \citet{sawai+13} showed for 
the first time that a sub-magnetar-class magnetic field was amplified by the MRI to the magnetar-class strength in the highest resolution 
axisymmetric global simulation to date. Furthermore, \citet{sawai+14} found by combining the neutrino transport approximately 
that the angular momentum transport by the MRI-amplified magnetic field contributes to the enhancement of the neutrino heating and 
eventually leads to neutrino-driven explosion. 

The aim of this Letter is to comprehend, using a new semi-global model, three-dimensional (3D) evolution of the MRI in the collapsed-core 
with the global structure of MHD variables. To numerically resolve the MRI with the wavelength much shorter than the global scale, a thin 
layer approximation is adopted. The nonlinear evolution of the MRI in the stably stratified layer around the neutrinosphere and its 
dependence on the magnetic flux threading the core are studied quantitatively. The MRI-sustained turbulent heating and the suitable 
condition for it are finally discussed.  
%%%%%%%%%%%%%%%%%%%%%%%%%%%%%%%%%%%%%%%%%%%%%%%%%%%%%%%%%%%%%%%%%%%%
%%%%%%%%%%%%%%%%%------------S2 Simulation Setup---------%%%%%%%%%%%%%%%%%%%%%%%%%%%%%%%%%
%%%%%%%%%%%%%%%%%%%%%%%%%%%%%%%%%%%%%%%%%%%%%%%%%%%%%%%%%%%%%%%%%%%%
\section{Model Setup}
%%%%%%%%%%%%%%%%%%%%%%%%%%%%%%%%%%%%%%%%%%%%%%%%%%%%%%%%%%%%%%%%%%%%
\begin{figure}[tp]
\scalebox{0.51}{{\includegraphics{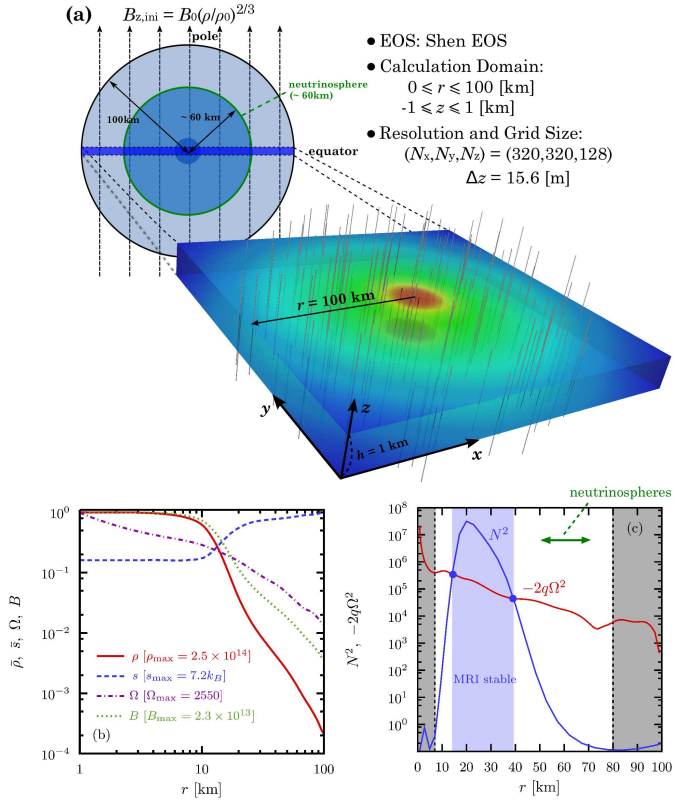}}} 
\caption{(a) Simulation setup. (b) Initial profiles of density (red solid), entropy (blue dashed), angular velocity (purple dash-dotted) and 
magnetic field (greed dotted) in our model. (c) Stability of the system to the MRI. The blue and red lines denote the LHS and 
RHS of equation (7). The positions of neutrinospheres are indicated by the horizontal arrow.}
\label{fig1}
\end{figure}
%%%%%%%%%%%%%%%%%%%%%%%%%%%%%%%%%%%%%%%%%%%%%%%%%%%%%%%%%%%%%%%%%%%%
To resolve the MRI-unstable modes, we focus only on the equatorial region of the core and adopt a thin layer approximation with $r \gg h$, 
where $r$ and $h$ are the radius and height of the system. In Figure 1(a), our semi-global model is shown schematically. Under this 
approximation, the vertical gravity is neglected and thus the physical variables are homogeneous initially in the vertical direction. The 
computational domain covers $0 \le r \le 100\ {\rm [km]}$ in the radius, $0 \le \phi \le 2\pi $ in the azimuth, and $-1 \le z \le 1\ {\rm [km]}$ in 
the height. The cylindrical structure of the thin-layer approximated supernova core is mapped onto the Cartesian domain with 
$L_x = L_y = 200\ {\rm km}$ and $L_z = 2\ {\rm km}$ to avoid the coordinate singularity. 

The fundamental equations are compressible MHD equations written by 
\begin{eqnarray}
\frac{\partial \rho}{\partial t} & = & - \nabla\cdot (\rho \bm{u})  \;, \label{eq1} \\ 
\frac{{D }\bm{u}}{ D t}  & = & - \frac{\nabla P}{\rho}  + \frac{\bm{J}\times\bm{B}}{c\rho} + \frac{\nabla \cdot (2\rho \nu_0 {\bm S})}{\rho} 
+ \bm{g}  \;, \ \ \ \ \label{eq2} \\
\frac{D\epsilon }{D t} & = & - \frac{P\nabla\cdot \bm{u}}{\rho} 
+ 2\nu_0 \bm{S}^2 + \frac{4\pi \eta_0\bm{J}^2}{\rho c^2}  \;, \label{eq3} \\
\frac{\partial \bm{B} }{\partial t} & = & \nabla \times (\bm{u} \times \bm{B} - \frac{4\pi \eta_0 \bm{J}}{c})\;, \label{eq4}
\end{eqnarray}
with
\begin{equation}
\bm{J}  =  \frac{c \nabla \times \bm{B}}{4\pi} \;, S_{ij}   =  \frac{1}{2}\left( \frac{\partial u_i}{\partial x_j} 
+  \frac{\partial u_j}{\partial x_i} - \frac{2}{3}\delta_{ij}\frac{\partial u_i}{\partial x_i} \right) \;, 
\end{equation}
where $\epsilon$ is the specific internal energy, $\bm{J}$ is the current density, and $S_{ij}$ is the strain rate tensor. The viscosity and 
magnetic diffusivity are represented by $\nu_0$ and $\eta_0$, respectively. We employ the equation of state based on the relativistic 
mean-field theory (Shen et al. 1998). 

We set an initial equilibrium model based on a post-bounce core from an axisymmetric hydrodynamic simulation of rotating core-collapse 
\citep[see][for the method]{takiwaki+14}. About $100$ ms after the core bounce, the shock wave has reached $\sim 200$ km, and the 
post-shocked region is settled into a quasi-hydrostatic structure. The radial distribution of the hydrodynamic variable along the equator 
of the simulation is extracted, and the stratified supernova core is reconstructed within our model. By assuming the gravity force balancing 
with the pressure gradient and centrifugal forces, a cylindrically symmetric hydrostatic structure is retained. 

The radial profiles of the density, entropy, and angular velocity adopted as the initial setting are shown in Figure 1(b) by red solid, blue 
dashed, and purple dash-dotted lines. They are normalized by their maximum values ($\rho_{\rm max} = 2.5\times 10^{14}\ {\rm g/cm^3}$, 
$s_{\rm max} = 7.2 k_B$ and $\Omega_{\rm max} = 2500\ {\rm rad/sec}$). The angular velocity adopted here corresponds to that of a 
rapidly rotating PNS \citep[e.g.,][]{ott+06}. The initial configuration of the magnetic field is given by 
\begin{equation}
B_z (r) = B_0 [\rho(r)/\rho_0]^{2/3} \;, 
\end{equation}
where $\rho_0$ is the density at the reference point $r = 30\ {\rm km}$ and $B_0$ is the magnetic field there. The dependence of the 
MRI-driven turbulence on $B_0$ is studied in \S~3.2. 

In our simulation, the profile of the electron fraction $Y_e$ is assumed to be constant with time from the initial value 
because neutrino transport is not solved. The convective stability is thus determined only by the entropy gradient. Since the effect of 
the $Y_e$-gradient on the MRI is much weaker than that of the entropy gradient in the region of our interest 
($8 {\rm km} \lesssim r \lesssim 80 {\rm km}$), we anticipate that, at least, the conclusion of this work is independent from it.

The stability condition of the stratified system to the MRI is written by \citep[e.g.,][]{balbus+94,masada+06}
\begin{equation}
N^2 > - 2q\Omega^2  \;,
\end{equation}
with 
\begin{equation}
N^2 \equiv  -\frac{1}{\gamma\rho} \nabla P \cdot \nabla \ln (P\rho^{-\gamma})\;, \  \ \ q = \frac{\partial \ln \Omega}{\partial \ln r} \;,
\end{equation}
where $\Omega$ is the angular velocity and $q$ is the shear rate. The LHS is the squared Brunt-V\"ais\"ala frequency due to the 
entropy gradient and the RHS is the destabilization effect by the MRI. In Figure 1(c), the LHS and 
RHS of equation (7) evaluated from the initial profiles are shown as a function of the radius. 
Our model comprises three characteristic regions: inner and outer MRI-unstable regions ($r \lesssim 15\ {\rm km}$ 
and $40\ {\rm km} \lesssim r$), and  middle MRI-stable region ($15\ {\rm km} \lesssim r \lesssim 40\ {\rm km}$). 
The horizontal arrow indicates the positions of (energy-integrated) neutrinospheres, which is smallest for heavy 
neutrinos ($\sim 50$ km) and largest ($\sim 70$ km) for electron neutrinos. The 
neutrinospheres are located in the MRI-unstable region initially. 

All the variables are assumed to be periodic in the vertical direction, whereas stress-free and perfect conductor boundary conditions are 
used in the horizontal direction for the velocity and magnetic fields. To reduce the artifacts due to the Cartesian geometry, 
the profile of the angular velocity is fixed in buffer regions denoted by gray shade in Figure 1(c) ($r < 8\ {\rm km}$ and $r > 80\ {\rm km}$). 
Furthermore, we assume $100$ times larger magnetic diffusivity in the buffer region than that in the region of $8\ {\rm km} \le r \le 80\ {\rm km}$. 
The MRI-driven turbulence is thus artificially quenched in the buffer region. 

The fundamental equations are solved by the second-order Godunov-type finite-difference scheme which employs an approximate MHD 
Riemann solver \citep{sano+98,masada+12}. The magnetic field is evolved with CMoC-CT method \citep{evans+88,clarke96}. 
The viscosity and resistivity are chosen as $\nu_0 = \eta_0 = 10^{12}\ {\rm cm^2/sec}$ in the buffer region, and set to zero in the 
other region. The spatial resolution of ($N_x, N_y, N_z$) 
$=$ $(320,320,128)$ is adopted for all the simulation runs. The grid convergence of our model will be studied 
carefully in a subsequent paper. A small ($1$\%) random perturbation relative to the unperturbed initial velocity 
field is added when the calculation starts. 
%%%%%%%%%%%%%%%%%%%%%%%%%%%%%%%%%%%%%%%%%%%%%%%%%%%%%%%%%%%%%%%%%%%%
%%%%%%%%%%%%%%%%%------------S3 Simulation Results---------%%%%%%%%%%%%%%%%%%%%%%%%%%%%%%%%
%%%%%%%%%%%%%%%%%%%%%%%%%%%%%%%%%%%%%%%%%%%%%%%%%%%%%%%%%%%%%%%%%%%%
\section{Simulation Results}
%%%%%%%%%%%%%%%%%%%%%%%%%%%%%%%%%%%%%%%%%%%%%%%%%%%%%%%%%%%%%%%%%%%%
\subsection{Fiducial Run with $B_0 = 10^{12}\ {\rm G}$}
%%%%%%%%%%%%%%%%%%%%%%%%%%%%%%%%%%%%%%%%%%%%%%%%%%%%%%%%%%%%%%%%%%%%
\begin{figure}[tbp]
\scalebox{0.25}{{\includegraphics{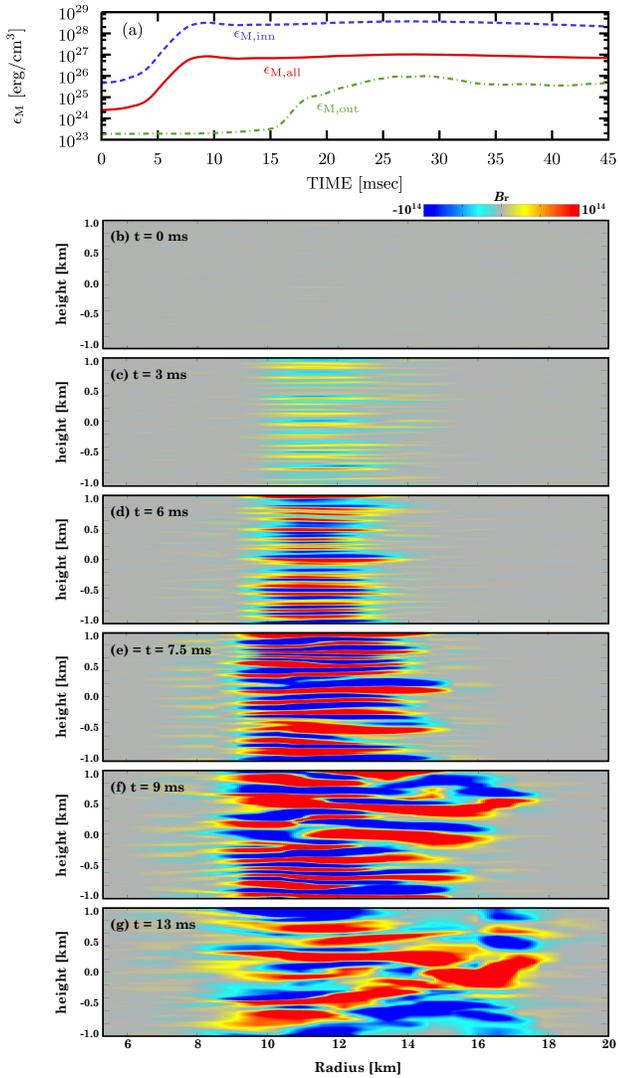}}} 
\caption{(a) Time series of volume-averaged magnetic energy [entire volume (red solid), inner volume (blue dashed), and outer 
volume (green dash-dotted) averages]. (b)-(f) Snapshots of $B_r$ in the inner MRI-unstable region on the $r$--$z$ cutting plane. }
\label{fig2}
\end{figure}
%%%%%%%%%%%%%%%%%%%%%%%%%%%%%%%%%%%%%%%%%%%%%%%%%%%%%%%%%%%%%%%%%%%%
\begin{figure*}[tbp]
\begin{center}
\scalebox{0.6}{{\includegraphics{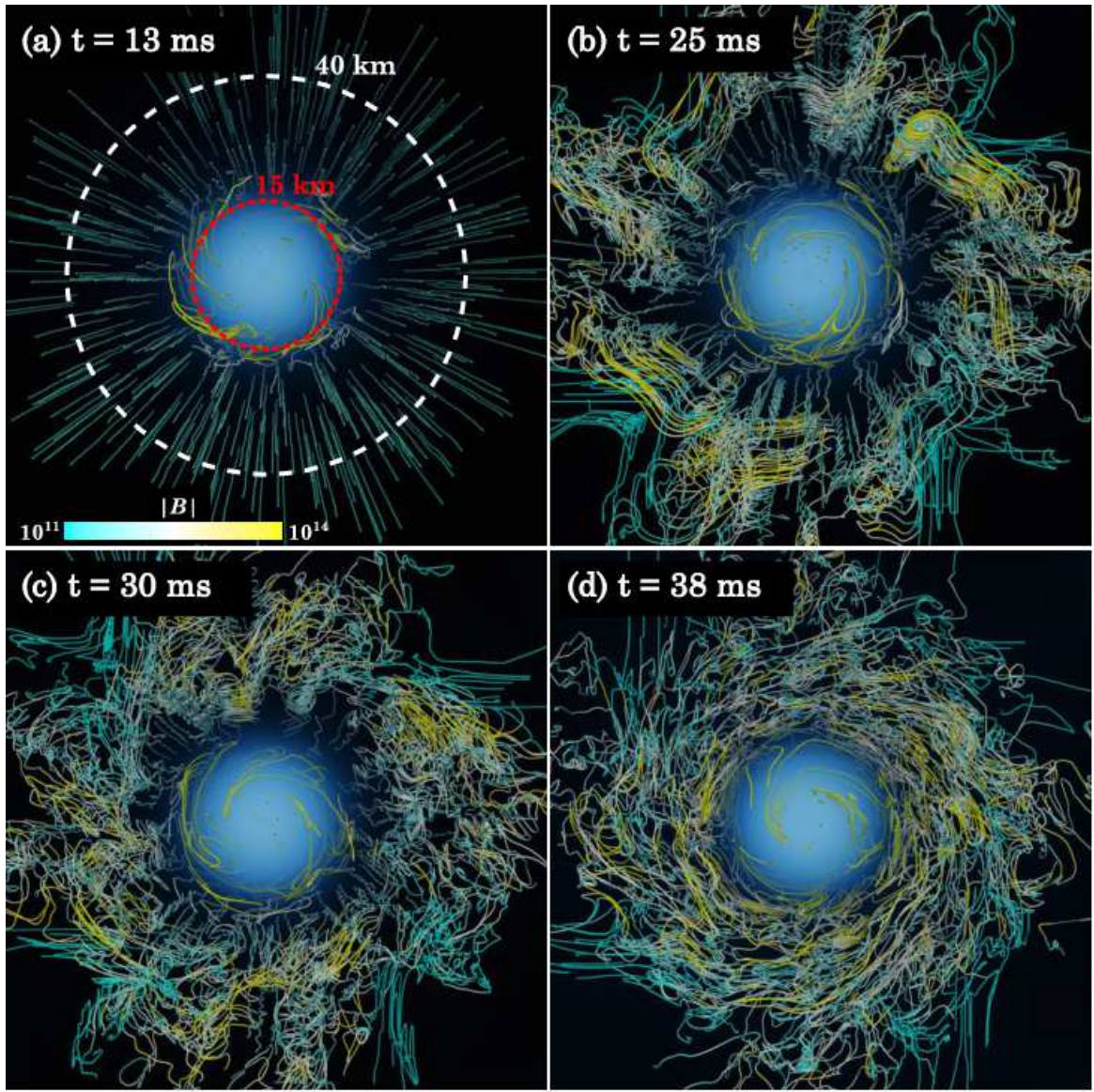}}} 
\caption{3D visualization of the magnetic field line. Panels (a)--(d) correspond 
to the sequential snapshots when $t = 13$--$38\ {\rm ms}$. The magnetic field line with the color denoting its absolute strength is 
superimposed on the volume visualization of the density. }
\label{fig3}
\end{center}
\end{figure*}
%%%%%%%%%%%%%%%%%%%%%%%%%%%%%%%%%%%%%%%%%%%%%%%%%%%%%%%%%%%%%%%%%%%%
The temporal-evolution of the MRI in a fiducial model with $B_0 = 10^{12}\ {\rm G}$ is elaborated. Shown in Figure 2 (a) is the time series 
of volume-averaged magnetic energy. The red solid, blue dashed and green dash-dotted lines denote the entire volume, inner volume 
($8\ {\rm km} \le r \le 15\ {\rm km}$), and outer volume ($15\ {\rm km} \le r \le 80\ {\rm km}$) averages, respectively. After the initial 
exponential growth phase, the magnetic energy is saturated at $t \simeq 10\ {\rm ms}$. By comparing the inner and outer magnetic 
energies, it is found that the early evolution is dominated by the MRI growth in the inner region. The delay of the MRI growth in 
the outer MRI-unstable region would be because our simulation cannot sufficiently capture the most unstable MRI wavelength there.

Panels (b)--(f) in Figure 2 present the snapshots of the radial magnetic field ($B_r$) in the inner MRI-unstable region on the $r$--$z$ 
cutting plane at $t = 0$--$12.5\ {\rm ms}$, with the red (blue) tone denoting the positive (negative) $B_r$. As seen in earlier studies, the 
evolution and disruption of the channel structure are observed \citep[e.g.,][]{obergaulinger+09,masada+12}. Since the convective motion 
does not develop in the stably stratified region, the secondary instabilities, such as parasitic instability, are responsible for the disruption 
of the channels \citep[see, e.g.,][for the effects of the secondary instabilities and background flow motion on the MRI]{obergaulinger+09, sawai+13}. 

The growth of the MRI in the outer region becomes prominent after $t \simeq 15\ {\rm ms}$ [green dash-dotted line in Figure 2(a)]. 
Since the magnetic energy stored in the outer region is small compared to the inner region, it gives a minor contribution to the total 
magnetic energy. However, the material mixing by the MRI-driven turbulence plays a crucial role in reducing the entropy gradient 
and thus in developing the fully-turbulent PNS surface. 

Figure 3 is the 3D visualization of the magnetic field line in the late evolutionary phase after $t \simeq 10\ {\rm ms}$ (view from the 
$z$-direction). Panels (a)--(d) correspond to the sequential snapshots. The magnetic field line with the color denoting its absolute strength 
is superimposed on the volume visualization of the density. The red and white dashed lines in the panel (a) are the circles with $r = 15$ 
and $40\ {\rm km}$ as a reference. 

In the late evolutionary stage, the MRI begins to grow in the outer MRI-unstable region ($r \gtrsim 40\ {\rm km}$) at around $t \simeq 15\ {\rm ms}$ 
[panel (a)]. Intriguingly, as time advances, the MRI-turbulent region spreads radially-inward [panels (b) \& (c)], and 
penetrates into the middle MRI-stable region [panel (d)]. The turbulent mixing which can flatten the entropy gradient would be responsible 
for it. As a result, whole the convectively stable region around the neutrinosphere is overwhelmed with the MHD turbulence. 

The multistage evolution of the MRI can be seen also in the energy spectra. In Figure 4, the Fourier spectra of the 
magnetic energy, averaged over the azimuth and a given time span at (a) $r = 10\ {\rm km}$ (inner MRI-unstable region), 
(b) $r = 30\ {\rm km}$ (middle MRI-stable region), and (c) $r = 60\ {\rm km}$ (outer MRI-unstable region), are shown. The horizontal axis 
represents the vertical wavenumber $k$ normalized by $k_c \equiv 2\pi/L_z$. The different lines correspond to different time spans. The 
black dashed line is the reference slope of $\propto k^{-5/3}$. 

The magnetic energy inversely cascades from the small scale to the larger scale as time passes. In the inner MRI-unstable region 
[panel (a)], it begins to evolve immediately after the calculation starts and a saturated MRI-turbulent state is achieved at 
$t \simeq 10\ {\rm ms}$. After $t \simeq 10\ {\rm ms}$, the outer MRI-unstable region begins to be destabilized [panel (c)]. 
%In this region, somewhat surprisingly, the magnetic energy is amplified to $10^{10}$ times the initial one at the saturated stage. 
The middle MRI-stable region is subsequently destabilized after $t \simeq 20\ {\rm ms}$ and is finally settled into the turbulent state at 
$t \simeq 40\ {\rm ms}$ [panel (b)]. 
%%%%%%%%%%%%%%%%%%%%%%%%%%%%%%%%%%%%%%%%%%%%%%%%%%%%%%%%%%%%%%%%%%%%
\subsection{Dependence on $B_0$}
%%%%%%%%%%%%%%%%%%%%%%%%%%%%%%%%%%%%%%%%%%%%%%%%%%%%%%%%%%%%%%%%%%%%
\begin{figure*}[htbp]
\scalebox{0.43}{{\includegraphics{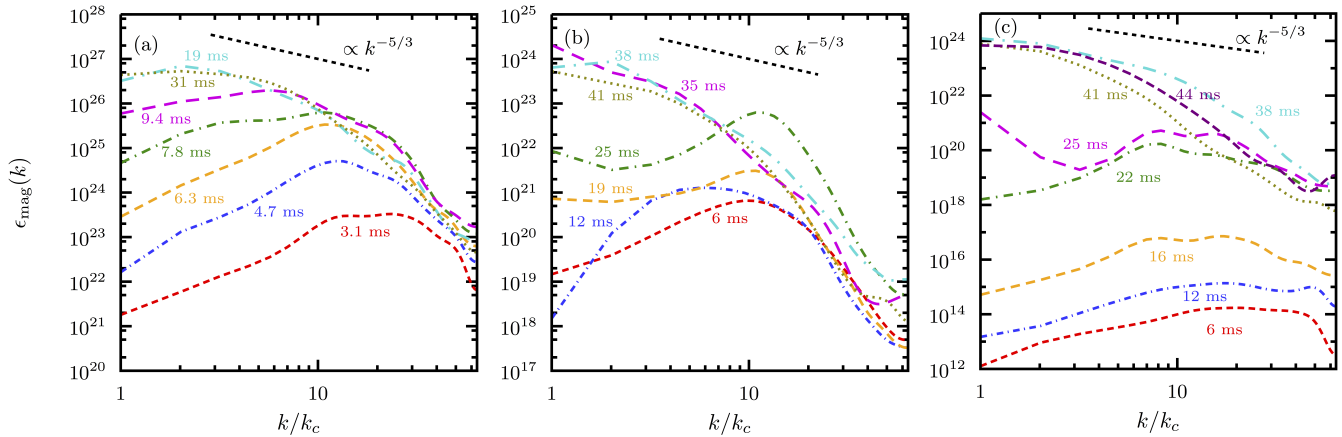}}} 
\caption{Fourier spectra of the magnetic energy at (a) $r = 10\ {\rm km}$, (b) $r = 30\ {\rm km}$, 
and (c) $r = 60\ {\rm km}$, which are averaged over the azimuth and a given time span. The horizontal axis 
is the vertical wavenumber $k$ normalized by $k_c \equiv 2\pi/L_z$. The black dashed line is the reference slope of $\propto k^{-5/3}$. }
\label{fig4}
\end{figure*}
%%%%%%%%%%%%%%%%%%%%%%%%%%%%%%%%%%%%%%%%%%%%%%%%%%%%%%%%%%%%%%%%%%%%
The property of the saturated MRI-driven turbulence is examined by varying $B_0 $ from $10^{11} $ to $5\times 10^{13}\ {\rm G}$ without 
changing the initial equilibrium model. 

Figure 5 (a) shows the $B_0$-dependence of the mean magnetic energy density $\langle\langle \epsilon_{M} \rangle\rangle$, where the single 
angular brackets denote the volume average spanning in the range of $r_{\rm in} \le r \le r_{\rm out}$ and the additional 
angular brackets denote the time average in the range of $t_s \le t \le t_e$ [$r_{\rm in}$ ($r_{\rm out}$) is $10\ {\rm km}$ ($80\ {\rm km}$) and 
$t_s$ ($t_e$) is $30\ {\rm ms}$ ($40\ {\rm ms}$) for all the models]. 
The red circle represents the MRI-turbulent model and the green cross is the model in which the MRI does not grow.  
The orange dotted line is the reference slope with $\propto B_0^{2}$. The expected magnetic energy stored in the full spherical shell of 
$r_{\rm in} \le r \le r_{\rm out}$, which is defined by 
\begin{equation}
E_{\rm mag} = \frac{4}{3}\pi(r_{\rm out}^3 - r_{\rm in}^3) \langle\langle \epsilon_{M} \rangle\rangle \;,
\end{equation}
is presented in the right axis just for the reference. 

In the range $B_0 \ll 10^{13}\ {\rm G}$, $\langle\langle \epsilon_{M} \rangle\rangle$ is proportional to $B_0^{2}$. This is similar to 
the recent studies of the MRI in the stratified accretion disk (e.g., Suzuki et al. 2010). 
%, though  the radial global structures of the MHD variables are considered in our supernova model.  
The dependency becomes weaker with the increase of $B_0$ and $\langle\langle \epsilon_{M} \rangle\rangle $ 
hits the ceiling at $\epsilon_{\rm upper} \simeq 4\times 10^{27}\ {\rm erg/cm^3}$ in the range $B_0 \gtrsim 10^{13}\ {\rm G}$. 
Except for the most strongly magnetized model in which the MRI does not grow, whole the convectively stable region around the 
neutrinosphere is overwhelmed with the MHD turbulence. %after the multistage evolution of the MRI. 

The upper threshold $\epsilon_{\rm upper}$ seems to be determined by the free energy stored in the differential rotation. 
By taking the initial rotation profile, the volume-averaged shear energy $\langle \epsilon_{\rm shear} \rangle$, which can be tapped for 
the amplification of the magnetic field, is evaluated as 
\begin{equation}
\langle \epsilon_{\rm shear} \rangle  \equiv  \left\langle \frac{1}{2}\left( \frac{\Delta\Omega}{\Omega} \right)^2 \rho v_{\phi}^2 \right\rangle 
 \simeq  4 \times 10^{27} \ {\rm erg/cm^3}  \;,
\end{equation}
where $\Delta \Omega \equiv (\partial \Omega/\partial r)\ {\rm d}r $ is the rotational shear \citep[see, e.g.,][]{spruit08}. $\langle \epsilon_{\rm shear} \rangle$ is 
presented by the blue dashed line in the panel (a). There is a good agreement between $\epsilon_{\rm upper}$ and $\langle \epsilon_{\rm shear}\rangle $. 
This suggests that the convectively-stable layer around the neutrinosphere becomes fully-turbulent when the condition 
\begin{equation}
\langle \epsilon_{M, {\rm init}} \rangle \lesssim \langle \epsilon_{\rm shear} \rangle \;, 
\end{equation}
is fulfilled, where $\epsilon_{M,{\rm init}}$ is the initial magnetic energy density. 
The most strongly magnetized model does not meet this precondition. This would be the reason why the MRI-turbulence does not grow in that model.  

%To verify the presence of the upper threshold for the MRI growth, we additionally simulate the strongly magnetized model with 
%$B_0 = 5\times 10^{13}\ {\rm G}$, which has an initial volume-averaged $\epsilon_{M}$ larger than $\epsilon_{\rm shear}$. 
%The $\langle \epsilon_{M} \rangle $ of this model is shown by a green cross in Figure 5 (a). 
%It is found that the MRI does not grow in this model and thus the magnetic energy remains almost steady as time passes. 
%This suggests that the fully-turbulent PNS can form only when the condition 
%\begin{equation}
%\int_{r_{\rm in}}^{r_{\rm out}} \epsilon_{M} {\rm d}V \lesssim \epsilon_{\rm shear} \;, 
%\end{equation}
%is satisfied at the post-bounce core. 

%There is an another possibility that the upper threshold of the magnetic energy is due to the geometrical effect because the box height puts a restriction 
%on the MRI wavelength. The larger box height allows the longer MRI wavelength being activated, providing the larger magnetic energy. To examine this 
%possibility, the model with $4L_z$ and $B_0 = 10^{13}\ {\rm G}$ is simulated without changing the resolution in the vertical direction. $\langle \epsilon_{M} \rangle$ 
%of this model is shown by green open circle in the Figure 5 (a). We can find that the saturated magnetic energy is little dependent on the 
%box height. We can thus conclude that the upper threshold is determined by the free energy stored in the differential rotation. 

The upper threshold for the saturated magnetic energy is a main difference between the MRI in CCSNe and that in the 
accretion disk. In the supernova core, the force balance is mainly achieved between the gravity and the pressure gradient 
force unlike the accretion disk in which the gravity is balanced with the centrifugal force. Since the free energy stored in the differential rotation 
is much lower than the thermal energy in the supernova core, the saturated magnetic energy is constrained predominantly by the shear 
energy rather than by the thermal energy. 

The MRI-driven turbulence sustains the turbulent heating around the neutrinosphere. Figure 5 (b) shows the $B_0$-dependence 
of the MRI-luminosity $\mathcal{L}_{\rm MRI} $, which is equivalent to the turbulent heating rate and is defined by 
\begin{equation}
\mathcal{L}_{\rm MRI}   =   \int_{r_{\rm in}}^{r_{\rm out}}  \int_{t_s}^{t_e}4\pi r^2 w_{\rm M} q \Omega\ {\rm d}r {\rm d}t\;, \\
\end{equation}
where $w_{\rm M} \equiv - B_r B_\phi /4\pi $ is the turbulent Maxwell stress \citep[see][]{thompson+05,masada+12}. 
Note that the spherical symmetry is assumed here.  
The red circle denotes the model dominated by the MRI-driven turbulence, and the green cross is the model in which the MRI does not grow. 

$\mathcal{L}_{\rm MRI} $ is increased with $B_0$ in the regime $B_0 \lesssim 10^{13}\ {\rm G}$. However, in the most strongly magnetized model, 
the turbulent heating is drastically suppressed because the MRI-driven turbulence does not develop. 
This suggests that, in the rapidly rotating PNS assumed in this work, the pulsar-sized relatively weak poloidal magnetic field 
($10^{12}$--$10^{13}\ {\rm G}$) is the most suitable for the MRI-sustained turbulent heating and then provides 
$\mathcal{L}_{\rm MRI} \simeq 10^{51}\ {\rm erg/sec}$ in the supernova core. 

By extrapolating the simulation result of the rapidly rotating PNS, we finally evaluate $\mathcal{L}_{\rm MRI}$ expected in 
PNSs with extremely rapid rotation. When assuming $w_{\rm M} \propto \Omega^2$, $\mathcal{L}_{\rm MRI}$ 
should be proportional to $\Omega^3 $ from the equation (12). 
In addition, equation (11) suggests that the maximum value of $B_0$ for enabling the MRI-driven turbulence is proportional to $\Omega$. 
According to \citet{burrows+07} and \citet{takiwaki+09}, the most rapidly rotating PNS is expected to have $3$ times higher $\Omega_0$ 
than that of our simulation model ($\Omega_0 = 2500\ {\rm rad/s}$). To infer an upper bound of $L_{\rm MRI}$,  we plot the blue-dashed 
line in Figure 5 (b), which corresponds to the extrapolated model. The MRI-luminosity is sensitive to $\Omega$ and would have larger 
impact on the supernova dynamics in the faster spinning PNSs. 

%%%%%%%%%%%%%%%%%%%%%%%%%%%%%%%%%%%%%%%%%%%%%%%%%%%%%%%%%%%%%%%%%%%%
\begin{figure*}[tbp]
\scalebox{0.78}{{\includegraphics{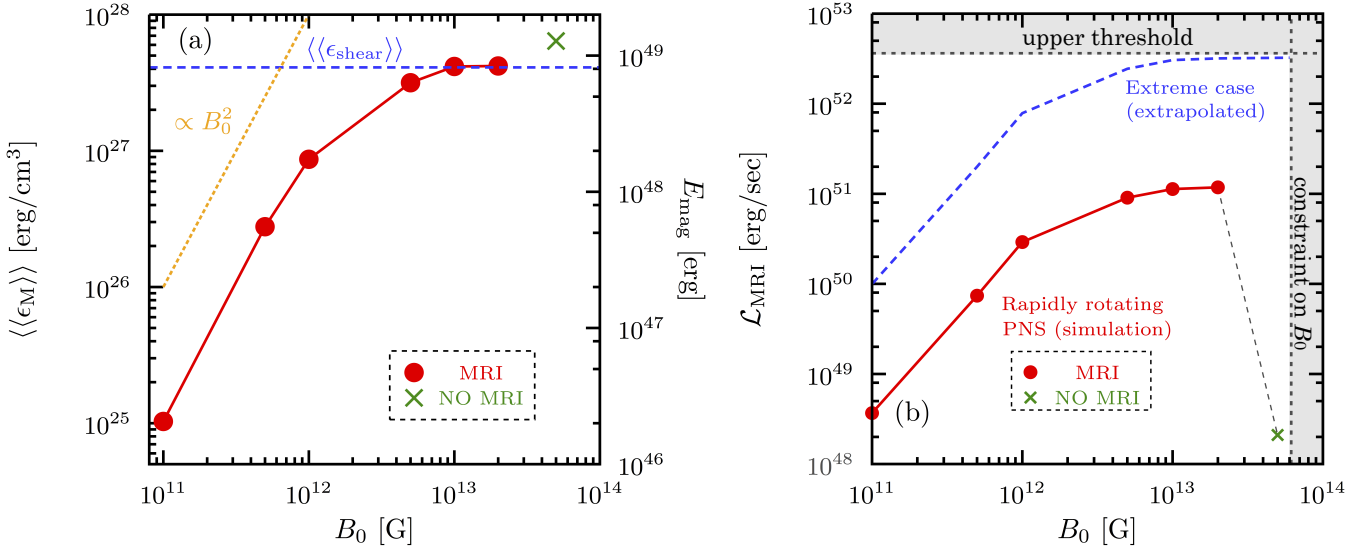}}} 
\caption{(a) $\langle\langle \epsilon_{M} \rangle\rangle$ as a function of $B_0$. 
The orange dotted and blue dashed lines denote the reference slope with $\propto B_0^{2}$ and the upper threshold $\epsilon_{\rm shear}$. 
(b) $B_0$-dependence of $\mathcal{L}_{\rm MRI} $ (red circles and green cross). The blue dashed line denotes $\mathcal{L}_{\rm MRI} $ 
expected in the PNS with $3\Omega_0$. The gray shades are constraint on $B_0$ and $L_{\rm MRI}$ for the extreme case.}
\label{fig5}
\end{figure*}
%%%%%%%%%%%%%%%%%%%%%%%%%%%%%%%%%%%%%%%%%%%%%%%%%%%%%%%%%%%%%%%%%%%%
%%%%%%%%%%%%%%%%%------------S5 Discussion & Summary----------%%%%%%%%%%%%%%%%%%%%%%%%%%%%%
%%%%%%%%%%%%%%%%%%%%%%%%%%%%%%%%%%%%%%%%%%%%%%%%%%%%%%%%%%%%%%%%%%%%
\section{Summary}
%%%%%%%%%%%%%%%%%%%%%%%%%%%%%%%%%%%%%%%%%%%%%%%%%%%%%%%%%%%%%%%%%%%%
In this Letter, we studied the 3D evolution of the MRI by a semi-global model simulating the supernova core. The nonlinear property of 
the MRI-driven turbulence in the stably stratified layer around the neutrinosphere and its dependence on the initial magnetic flux were 
quantitatively elucidated. 

It was found that the MRI-driven turbulence excited in inner and outer MRI-unstable regions gradually erodes the strongly-stratified 
middle MRI-stable region. Finally, whole the convectively stable region around the neutrinosphere was overwhelmed with the MHD 
turbulence. The intensity of the saturated MRI-driven turbulence in the PNS was affected by the magnetic flux threading the collapsed core: 
while the magnetic energy sustained by the MRI-driven turbulence was proportional to $B_0^{2}$ in the range $B_0 \ll 10^{13}\ {\rm G}$, 
it is limited by the free energy stored in the differential rotation when $B_0 \gtrsim 10^{13}\ {\rm G}$. 

%The larger height of the simulation box allows the longer MRI wavelength being activated, providing the larger magnetic energy. There is 
%thus a possibility that the box height puts a restriction on the MRI wavelength and thus upper threshold of the magnetic energy found in \S~3.2. 
%To examine the effect of the box height, a test model with $4L_z$ and $B_0 = 10^{13}\ {\rm G}$ is simulated without changing the resolution 
%in the vertical direction, i.e., $N_z = 512$. The mean magnetic energy density of this model is shown by the orange open circle in the Figure 5 (a). 
%We can find that $\langle \epsilon_{M} \rangle $ is little dependent on the box height. This confirms that the upper threshold is not determined 
%by the box size, but by the free energy stored in the differential rotation. 

Our results indicate that the turbulent neutrinosphere is a natural consequence of rotating core-collapse even in the absence of convection. In addition, 
the MRI-sustained turbulent heating is the most effective when the pulsar-sized magnetic field ($10^{12}$--$10^{13}$ G) threads the post-collapse core 
in the rapidly rotating PNS. 

Recently, the effects of the turbulence on the supernova dynamics are getting a lot more attention in numerical 
modeling of CCSNe. Not only the turbulent heating discussed here, the angular momentum transport and the material mixing in the meridional plane 
due to the turbulence should also play a crucial role in facilitating CCSNe \citep[e.g.,][]{murphy+11,hanke+12,handy+14,sawai+14}. 
Aided by the detailed analysis of the MRI in the linear phase \citep[e.g.,][]{guillet+14},
future higher resolution 3D numerical modeling of CCSNe will uncover hidden properties of the MHD turbulence 
in the supernova core. We have attempted the very first step toward such models in this study. 

%%%%%%%%%%%%%%%%%%%%%%%%%%%%%%%%%%%%%%%%%%%%%%%%%%%%%%%%%%%%%%%%%%
\acknowledgments
We acknowledge the anonymous reviewer for constructive comments. Computations were carried on XC30 at NAOJ. 
This work was supported by the Grants-in-Aid for the Scientific Research from the 
Ministry of Education, Science and Culture of Japan (Nos. 24740125, 26870823, 24244036, 24103006, 26707013)
%%%%%%%%%%%%%%%%%%%%%%%%%%%%%%%%%%%%%%%%%%%%%%%%%%%%%%%%%%%%%%%%%%
%%%%%%%%%%%%%%%%%%%%%%%----------------- End of Body -----------------%%%%%%%%%%%%%%%%%%%%%%%%
%%%%%%%%%%%%%%%%%%%%%%%%%%%%%%%%%%%%%%%%%%%%%%%%%%%%%%%%%%%%%%%%%%

%%%%%%%%%%%%%%%%%%%%----------------- END OF PAPER -----------------%%%%%%%%%%%%%%%%%%%%
\end{document}